\documentclass[prd,letterpaper,tightenlines,nofootinbib,twocolumn,floatfix,showpacs]{revtex4}
\usepackage{amssymb,graphicx,amsmath,color,float,makecell}
\allowdisplaybreaks
\graphicspath{{./plotmaker/}}
\begin{document}
\title{A simple family of solutions of relativistic viscous hydrodynamics for fireballs with Hubble flow and ellipsoidal symmetry}
\author{M.~Csan\'ad$^{1}$, M.~I.~Nagy$^{1}$, Ze-Fang~Jiang$^{2,3,4}$ and T.~Cs\"org\H o$^{5,6}$}
\affiliation{$^1$ E{\"o}tv{\"o}s Lor{\'a}nd University, H-1117 Budapest, P{\'a}zm{\'a}ny P. s. 1/A, Hungary}
\affiliation{$^2$ Department of Physics and Electronic-information Engineering, Hubei Engineering University,~Xiaogan 432000,~China}
\affiliation{$^3$ Key Laboratory of Quark and Lepton Physics, Ministry of Education, Wuhan, 430079, China}
\affiliation{$^4$ Institute of Particle Physics, Central China Normal University, Wuhan 430079, China}
\affiliation{$^5$ Wigner RCP, Centre for Excellence of the Hungarian Academy of Sciences, H-1525 Budapest 114, P.O.Box 49, Hungary}
\affiliation{$^6$ EKU KRC, H-3200, Gy{\"o}ngy{\"o}s, M{\'a}trai \'ut 36, Hungary}

\begin{abstract}
New, analytic solutions of relativistic viscous hydrodynamics are presented, describing expanding fireballs
with Hubble-like velocity profile and ellipsoidal symmetry, similar to fireballs created in heavy ion collisions.
We find that with these specifications, one obtains solutions where the shear viscosity essentially does not
influence the time evolution of the system, thus these solutions are particularly adept tools to study the
effect of bulk viscosity alone, which always results in a slower decrease of energy density as well as
temperature compared to the case of perfect fluid. We investigate different scenarios for the bulk
viscosity and find qualitatively different effects on the time evolution which suggests that there is
a possibility to infer the value of bulk viscosity from energy density and temperature measurements in
high-energy heavy-ion collisions.
\end{abstract}

\date{\today}
\pacs{20.24, 20.25}
\maketitle

\section{Introduction}
Relativistic viscous hydrodynamics is a well suited tool to investigate the space-time evolution and transport properties of strongly interacting
quark-gluon plasma (QGP) produced in high energy heavy ion collisions~\cite{Bass:1998vz,Gyulassy:2004zy}. There has been tremendous progress in
studying the equations of relativistic hydrodynamics in recent years, in the perfect fluid as well as in the viscous
case~\cite{Heinz:2013th,Song:2010vis,Csand:2016arx,Ze-Fang:2017ppe,Pang:2018zzo,Wu:2018cpc,Jiang:2018wzu,Chen:2017zte}.

Analytic solutions, even with simple initial conditions, play an important role in understanding the properties of strongly coupled expanding
QCD matter. Some of the historically relevant exact analytic solutions (such as the Landau-Khalatnikov solution~\cite{Landau:1953gs}, Hwa-Bjorken
solution~\cite{Hwa:1974gn,Bjorken:1982qr}) gave much insight into the general features of expansion dynamics in high energy collisions.
Further, more recent solutions include rotating expanding solutions as well as a generalized equation of state. (We name a few such solutions,
in whose trail our present work fits: the Gubser solution~\cite{Gubser:2010ze,Gubser:2010ui,Hatta:2014gqa,Hatta:2014gga}, the CCHK
solution~\cite{Csorgo:2003rt,Csorgo:2003ry}, the CNC solution~\cite{Csorgo:2006ax,Csorgo:2008prc} and the CKCJ solutions~\cite{Csorgo:2018pxh,Csorgo:2018fbz}.)

While nowadays there are various analytic solutions at hand in the perfect fluid case, exact solutions of relativistic hydrodynamics which
take dissipative effects (viscosity, heat conduction) into account showed considerably slower progress (if at all). In part this is probably due to
the fact that the relativistic dissipative hydrodynamic equations are even more complex and involved than their non-dissipative counterparts.
A problem of possibly more fundamental nature arises from the fact that in the relativistic case even the correct form of the basic equations
is a topic not yet settled well enough. A simple approach is to take dissipative effects as first order corrections into account (such as the fluid
equations due to Eckart~\cite{Eckart:1940te} and Landau~\cite{Landau:VI:1959}). There are various second order equations corresponding to more
realistic physical scenarios (the most well-known among them being the Israel-Stewart theory~\cite{Israel:1979wp}).

Taking into account first order viscous corrections, the bulk viscosity causes locally isotropic deviations from perfect flow, as the bulk viscous
pressure creates a diagonal contribution to the stress tensor. Based on the results from AdS/CFT and lattice QCD calculations for bulk viscosity,
it was pointed out that the bulk viscosity contribution for a high-temperature QCD medium is negligible~\cite{Meyer:2007dy}. In contrast, the bulk
viscous contribution at low temperatures, especially at those close to the critical temperature, has an important correction effect~\cite{Kharzeev:2007wb}.

In this paper, new, mathematically exact and analytic solutions of relativistic viscous hydrodynamics are investigated.
The calculations in this paper include shear and bulk viscosity, and also heat conduction. However, we show that the shear viscosity as well as the
thermal conductivity effects cancel for the (Hubble-type) velocity profiles which the new solutions incorporate. We then investigate the temperature
and flow evolution from the presented solutions in order to study the effect of bulk viscosity.

\section{Basic equations and assumptions}

Let us first summarize the notations and the equations of viscous hydrodynamics we are using below. We work in flat space-time: the metric is Minkowskian
with the sign convention of $g_{\mu\nu} = diag(1,-1,-1,-1)$. The fluid motion is described by the four-velocity field $u^\mu$ (normalized to unity,
$u_\mu u^\mu = 1$) and the thermodynamic quantities: the pressure $p$, the energy density $\varepsilon$, the temperature $T$, the entropy density $s$.
Below we treat solutions also where there is a non-vanishing conserved particle number: in this case its density is denoted by $n$, and the corresponding
chemical potential by $\mu$. All these quantities are functions of the space-time coordinate $x$. We denote the dimension of the space with $d$; normally
$d{=}3$, but in some formulas it is worthwhile to remember how a particular number that is related to the dimensionality of space occurs. For example,
the trace of the Minkowski metric tensor $g_{\mu\nu}$ is ${g^{}_\mu}^\mu = \delta_\mu^\mu = d{+}1$.
(We use the Einstein index summation convention throughout.)

The equations of hydrodynamics are encompassed in the condition of energy and momentum conservation, which translates into the vanishing of the
four-divergence of the energy-momentum tensor $T_{\mu\nu}$:
\begin{align}
\partial_{\mu}T^{\mu\nu}=0.
\end{align}
When writing up the first order viscous corrections to the equations of hydrodynamics, one has to make a choice of the definition of the four velocity
(referred to as a choice of ''frame''). We now work in the Eckart frame: we treat the fluid velocity as that of the conserved particle number.
In this frame the form of the $T_{\mu\nu}$ tensor is
\begin{align}
T_{\mu\nu} &= (\varepsilon{+}p)u_\mu u_\nu {-} pg_{\mu\nu} {-} (g_{\mu\nu}{-}u_\mu u_\nu)\Pi \\
 &+ (q_\mu u_\nu {+} q_\nu u_\mu) + \pi_{\mu\nu}\nonumber
\end{align}
The first two terms are the same as in the case of ideal (ie. non-dissipative) hydrodynamics; the term with $\Pi$ describes bulk viscosity, the terms with the $q_\mu$ quantity
correspond to heat conduction, while the last term ($\pi_{\mu\nu}$) describes shear viscous effects.
The dissipative quantities $\Pi$, $q_\mu$ and $\pi_{\mu\nu}$ are subject to the conditions
\begin{align}
q_\mu u^\mu = 0,\qquad\pi_{\mu\nu} u^\nu = 0.
\end{align}
and in Navier-Stokes hydrodynamics, their explicit form (essentially uniquely determined from the requirement of the law of entropy increase) is
\begin{align}
q_\mu        &= \lambda(g_{\mu\nu}{-}u_\mu u_\nu)\big(\partial^\nu T - Tu^\rho\partial_\rho u^\nu\big),\label{e:NSq}\\
\Pi          &= - \zeta \partial_\rho u^\rho,\label{e:NSPi}\\
\pi_{\mu\nu} &= 2\eta \sigma^{\mu\nu},\label{e:NSpi}
\end{align}
where
\begin{align}
\sigma_{\mu\nu} &= \big[(g_{\mu\rho}{-}u_\mu u_\rho)\partial^\rho u_\nu + (g_{\nu\rho}{-}u_\nu u_\rho)\partial^\rho u_\mu\big] +\nonumber\\
             &\,\, - \frac{2}{d}(g_{\mu\nu}{-}u_\mu u_\nu)\partial_\rho u^\rho\label{e:sigmamunu}
\end{align}
is the shear stress stensor. In Israel-Stewart hydrodynamics, these fields are no longer given in the above form but are instead ``promoted'' to fields with their own dynamical equations:
\begin{align}
\tau_{_\Pi} u_\rho \partial^\rho \Pi &= -\zeta \partial^\rho u_\rho - \Pi,\label{e:IS1}\\
\tau_\pi u_\rho \partial^\rho \pi^{\mu\nu} &= 2\eta \sigma^{\mu\nu} - \pi^{\mu\nu},\label{e:IS2}
\end{align}
where $\tau_{_\Pi}$ and $\tau_\pi$ are the Israel-Stewart relaxation coefficients, and $\sigma^{\mu\nu}$ is given as above in Eq.~\eqref{e:sigmamunu}. The thermal conductivity $\lambda$, and the shear and bulk viscosity coefficients $\eta$ and $\zeta$ may depend on the thermodynamic quantities. In the solutions presented below, only the bulk viscosity plays an explicit role, while the shear viscosity and the heat conductivity effects cancel. For this reason we do not discuss the many assumptions on the thermal conductivity $\lambda$ here, but will return to its possible effects on our solutions after having worked them out systematically.

The bulk viscosity $\zeta$ for realistic strongly coupled QCD matter produced in heavy ion collisions is an open theoretical field
under investigation (see eg.~\cite{Karsch:2007jc,NoronhaHostler:2008ju,Harutyunyan:2018wdk,Ryu:2017qzn}).
Many actual model calculations (eg.~\cite{Arnold:2006fz}) show that $\zeta$ is usually negligible in compare to $\eta$. 
The current best estimates~\cite{Bernhard:2016tnd,Bernhard:2019bmu} indicate a peak of $\zeta/s$ around the transition temperature,
a feature consistent with theoretical models~\cite{Rougemont:2017tlu,Alqahtani:2017mhy}. However, in the solution presented below,
the terms proportional to $\eta$ vanish identically, so these solutions are well suited to study the effect of bulk viscosity alone.

In order to find exact analytic hydrodynamic solutions, one should proceed from simple assumptions toward more complicated ones. When attempting
to explore the effect of bulk viscosity, in our work we start by investigating two simple cases. One is the choice to assume $\zeta/s =$ const.,
ie.\@ treat the bulk viscosity as proportional to the entropy density: $\zeta = \zeta_0(s/s_0)$. The AdS/CFT approach conjectured strong coupling
limits for the specific shear and bulk viscosity~\cite{Benincasa:2005iv,Kovtun:2004de}: $\zeta\simeq\eta\left[\frac 1d-c_{s}^{2}\right]$, where
$c_s$ is the speed of sound\footnote{For the equation of state where $\varepsilon = \kappa p$ which we will use below, $c_s^2 = 1/\kappa$. Note
that for $\kappa = 3 = $const (ie. in the ultra-relativistic Boltzmann gas limit) the formula for the bulk viscosity gives zero. Indeed it is
known~\cite{Khalatnikov:1955} that the bulk viscosity vanishes for ultra-relativistic monatomic gases. This fact gives a confirmation for
the mentioned conjectured form of $\zeta$.}. In the AdS/CFT approach, a lower limit for the quantity $\eta/s$ (where $s$ is the entropy density)
was conjectured to be $\simeq1/(4\pi)$. The actual temperature dependence of $\eta/s$, however, is not well established; often the reasonable
assumption of $\eta/s=$ const. is made. Taking these together, it seems also reasonable to assume $\zeta/s =$ const.

An alternative simple assumption of ours is to take $\zeta = \zeta_0 =$ const., ie.\@ a bulk viscosity that remains constant along the time evolution.
This might seem oversimplified at first sight, but taking into account that in many model calculations (such as eg.~the one presented in
Ref.~\cite{Karsch:2007jc}), the $\zeta/s$ ratio actually increases with decreasing temperature (and thus in turn during the time evolution).
On the other hand, the $s$ entropy density decreases during time evolution, so it is not unreasonable to investigate the $\zeta =$ const.
case, where the increase of $\zeta/s$ is compensated by the decrease of $s$.

Finally, specifying the Equation of State (EoS) of the matter closes the set of equations.
In what follows, we consider the simple equation of state:
\begin{align}
\varepsilon=\kappa p,\qquad\kappa=\mathrm{const}.
\end{align}
This allows for simple, analytically and explicitly expressible solutions to be found. When applying such calculations to the description of experimental data, $\kappa$ is understood as an average EoS, but for our purposes here, a constant EoS is sufficient. Proceeding further, if there is a non-vanishing conserved density $n$, it obeys a continuity equation of the form
\begin{align}
\label{e:ncont}
\partial_{\mu}(nu^\mu)=0.
\end{align}
We investigate two distinct possibilities for the equation of state as well: a first case when there is a conserved particle number density $n$, which
obeys a continuity equation and with which the pressure can be expressed as $p = nT$. The second case is when there is no conserved particle number density,
and the energy density (though of as a thermodynamic potential function) is a function of the $s$ entropy density alone.
From simple thermodynamic identities one can compute the entropy density $s$ as a function of other variables in both cases:
\begin{align}
\varepsilon &= \kappa p,\quad p      = nT                     &&\rightarrow\; s = s_0\frac{n}{n_0} + n\ln\left(\frac{n_0}{n}\frac{T^\kappa}{T_0^\kappa}\right)
\label{e:entropy:n}\\
\varepsilon &= \kappa p,\quad \varepsilon\equiv\varepsilon(s) &&\rightarrow\; s = s_0\left(\frac{T}{T_0}\right)^\kappa.
\label{e:entropy:non}
\end{align}
Note the appearance of the constants $n_0$, $T_0$ and $s_0$ in the expressions of the entropy density. They arise from the fact that in classical
thermodynamics, entropy is only meaningful up to an arbitrary additive constant\footnote{Only quantum statistics, with a definite value of
$\hbar$ sets the entropy scale. For example, in the case of photon gas, which corresponds to second case in Eq.~(\ref{e:entropy:non}),
$\kappa{=}3$, and the constant is $s_0/T_0^3 = \frac{\kappa{+}1}{\kappa}\frac{4\sigma_{SB}}{c}$, with $c$ being the speed of light,
and $\sigma_{SB}$ the Stefan-Boltzmann constant. The latter contains the explicit value of $\hbar$.}. These constants were chosen in a way such that $s=s_0$ when $T=T_0$ (and $n=n_0$, if applicable). In line with this (and also
to preserve generality) we keep these free constants in the expressions of $s$, whenever needed.
Also note that the specific form of the continuity equation for $n$, Eq.~(\ref{e:ncont}) is characteristic to the Eckart frame: in this frame, the
fluid velocity is fixed to the current of the conserved particle number. Nevertheless, we may use the Eckart frame also in the case when we do not consider
a conserved particle number density.

The two cases for the equation of state, coupled together with the two possibilities for the $\zeta(s)$ dependence mentioned above, would give
four cases worthy of investigation. However, the case when there is conserved $n$ and $\zeta$ is proportional to $s$ deserves a slight
reconsideration. If the entropy density is expressed as in Eq.~(\ref{e:entropy:n}) above, the equations turn out to be overly complicated
in this case. Also, the assumption that $\zeta{\propto} s$ is (at least on the qualitative level) based on the results from AdS/CFT correspondence,
which is not very meaningful for the case of non-vanishing conserved $n$. So we omit this case as it is, and investigate two analogous cases instead.
Thus when there is conserved $n$ and $\zeta$ is not constant, we solve the equations with in two cases corresponding to additional assumptions.
One is motivated by the simple expression of the entropy $s$ in terms of $T$ in the case of vanishing $n$: we treat a case when in the expression
of $\zeta$ we write Eq.~(\ref{e:entropy:non}) as the expression of entropy in terms of temperature. The other reconsidered assumption might be
that in massless theories, $\zeta/s$ (and for the shear viscosity, $\eta/s$) plays the role of ''kinematic'' viscosity; the $\zeta/s=$ const.
condition would thus mean that this bulk kinematic viscosity is constant. In case of conserved particle number $n$, the similar assumption
might then be written as $\zeta\propto n$, instead of $s$.

In summary, the five different cases investigated below (exchanging the lastly mentioned two ones) are as follows:
\begin{align}
\bullet& \textnormal{ \textit{Case A:} No conserved $n$, and constant $\zeta$:}\nonumber\\
 &\quad \zeta = \zeta_0\textnormal{ (const)},\quad \varepsilon =\kappa p,\quad  p= p_0(T/T_0)^{\kappa+1}. \label{e:casea}\\
\bullet& \textnormal{ \textit{Case B:} With conserved $n$, and constant $\zeta$:}\nonumber\\
 &\quad \zeta = \zeta_0\textnormal{ (const)},\quad \varepsilon=\kappa p, \quad p = nT. \\
\bullet& \textnormal{ \textit{Case C:} No conserved $n$, and $\zeta{\propto} s$:}\nonumber\\
 &\quad \zeta = \zeta_0(T/T_0)^\kappa, \quad \varepsilon=\kappa p,\quad p= p_0(T/T_0)^{\kappa+1}. \\
\bullet& \textnormal{ \textit{Case D:} With conserved $n$, and $\zeta/n=$const:}\nonumber\\
 &\quad \zeta = \zeta_0(n/n_0), \quad \varepsilon=\kappa p,\quad p = nT.\label{e:cased}\\
\bullet& \textnormal{ \textit{Case E:} With conserved $n$, and ``$\zeta{\propto} s$'':}\nonumber\\
 &\quad \zeta = \zeta_0(T/T_0)^\kappa, \quad \varepsilon=\kappa p,\quad p = nT.\label{e:casee}\\
\bullet& \textnormal{ \textit{Case F:} No conserved $n$, and $\zeta{\propto} \Pi$:}\nonumber\\
 &\quad \zeta = \zeta_0(\Pi/\Pi_0), \quad \varepsilon=\kappa p,\quad p= p_0(T/T_0)^{\kappa+1}.\label{e:casef}
\end{align}
Note that in the cases when there is a conserved $n$, we can not assume $p= p_0(T/T_0)^{\kappa+1}$ to hold, since this would mean $s\propto n$, i.e. an adiabatic expansion, cf. Eq~(\ref{e:entropy:n}). Note also that Case F is only applicable in Israel-Stewart hydrodynamics, as in case of the Navier-Stokes equations, Eq.~\eqref{e:IS2} determines the connection between $\Pi$ and $\zeta$. Hence from now on we proceed as using Case F for Israel-Stewart hydrodynamics and Cases A-E for Navier-Stokes hydrodynamics.

In the following we investigate the simple Hubble-type relativistic flow and find solutions to the hydrodynamic equations with bulk viscosity,
separately in each cases specified above. In the following section, we enter into some details about how to find such solutions; the reader
who is interested in only the solutions themselves may skip directly to Section~\ref{s:solution}.

\section{Searching for Hubble-like solutions}\label{s:searching}

Let us now search for exact and analytic self-similar solutions of viscous hydrodynamics. Our starting point is the
Hubble-type ellipsoidal perfect fluid solutions of Refs.~\cite{Csorgo:2003rt,Csorgo:2003ry,Csanad:2004mm,Csanad:2012hr,Csanad:2014dpa}. While it is of academic interest to find exact solutions, we also note that Hubble flow was shown to develop in high energy heavy ion collisions~\cite{Chojnacki:2004ec} and also to be compatible with the data~\cite{Csanad:2009wc,Csanad:2014dpa}. The main
important combinations of the space-time coordinates are the proper time $\tau$, defined (inside the forward light-cone) as
\begin{align}
\tau = \sqrt{t^2{-} r_x^2 {-} r_y^2 {-} r_z^2}.
\end{align}
The other variable is the {\it scaling variable} $S$, which is defined with the three time-dependent principal axes $X(t)$, $Y(t)$ and $Z(t)$ of
an expanding ellipsoid as
\begin{align}
S=\frac{r_x^2}{X^2}+\frac{r_y^2}{Y^2}+\frac{r_z^2}{Z^2} =
 \frac 1{t^2}\left(\frac{r_x^2}{\dot X_0^2}+\frac{r_y^2}{\dot Y_0^2}+\frac{r_z^2}{\dot Z_0^2}\right),
\label{scal}
\end{align}
where the time evolution of the axes is given by
\begin{align}
X(t) = \dot X_0t,\qquad Y(t) = \dot Y_0t,\qquad Z(t) = \dot Z_0t.
\end{align}
The velocity field is chosen to be the simple Hubble-type flow, a three-dimensional generalization of Hwa-Bjorken-flow~\cite{Hwa:1974gn,Bjorken:1982qr}. Hubble-flow is interesting in itself: it is an asymptotic flow profile for any point-like explosion. In this case the flow-profile can be expressed as:
\begin{align}
u^{\mu}=\frac{x^\mu}{\tau}=\gamma\left(1,\frac{\dot X}{X}r_x,\frac{\dot Y}{Y}r_y,\frac{\dot Z}{Z}r_z \right).
\label{vel}
\end{align}
Here $\gamma$ is the Lorentz factor. Note that the time derivatives of the principal axes $X$, $Y$, $Z$ are constant in time.
It is easy to verify that the velocity field has vanishing acceleration, as well as that the co-moving derivative of the scaling variable $S$ is zero:
\begin{align}
u^\mu\partial_\mu S=0,\qquad u^\nu\partial_\nu u^\mu=0.
\label{d1}
\end{align}
It is easy to see that in any case when there is a conserved particle number density $n$, the solution for its continuity equation can be taken as the
same as in solutions for perfect fluid motion, see eg. Refs.~\cite{Csorgo:2003rt,Csorgo:2003ry,Csanad:2004mm,Csanad:2012hr}:
\begin{align}
n = n_0\left(\frac{\tau_0}{\tau}\right)^d\mathcal V(S),
\label{e:nsol}
\end{align}
with an arbitrary $\mathcal V(S)$ function of the scaling variable.

The time dependence of the other thermodynamic quantities will be influenced by the bulk viscosity $\zeta$. To proceed, one evaluates the components of the energy-momentum tensor. Because of the special nature of the given velocity field, both the perfect fluid terms and those describing viscosity turn out to be quite simple. The resulting expression is in the Navier-Stokes case (i.e. where Eq.~\eqref{e:NSPi} is true)
\begin{align}
T_{\mu\nu}&=\varepsilon\frac{x_\mu x_\nu}{\tau^2} - \Big(p {-} \zeta\frac d\tau\Big)\cdot\Big(g_{\mu\nu} {-} \frac{x_\mu x_\nu}{\tau^2}\Big)
+\nonumber\\
%q_\mu u_\nu{+}q_\nu u_\mu, &q_\mu u_\nu {+} q_\nu u_\mu =
&+\lambda\partial_\rho T\Big[
  \frac{x_\mu}{\tau}\Big(\delta^\rho_\nu{-}\frac{x_\nu x^\rho}{\tau^2}\Big) +
  (\mu{\leftrightarrow}\nu)\Big].
\label{e:Tmunu:expr}
\end{align}
Here again $d$ is the dimensionality of space, $d{=}3$. Note that for Hubble-flow, $\sigma^{\mu\nu}=0$ identically, hence (as mentioned above) all terms of $\pi_{\mu\nu}$ that would contain the
shear viscosity $\eta$ indeed cancel in the Navier-Stokes case.\footnote{Let us also note that in case of a true 1+1 dimensional flow, where $d=1$ and $u^\mu = (u^0,u^1)$, the shear viscosity part of the tensor $\pi^{\mu\nu}$ identically vanishes for \emph{any} $u^\mu$ field, not just for the Hubble flow specified here.}

Turning to the terms of $T_{\mu\nu}$ in Eq.~(\ref{e:Tmunu:expr}) that contain the thermal conductivity $\lambda$, we need to make some assumption
on $\lambda$ as well. To keep the investigation of the bulk pressure as simple as possible, from now on we will neglect
the terms that describe thermal conductivity, ie. take $\lambda{=}0$. However, it will turn out that some solutions that we find will have the
property that they remain valid solutions even for arbitrary $\lambda{\neq}0$.
We will come back to this when discussing the actual solutions.

Invoking the $\varepsilon = \kappa p$ equality, the equation to be solved turns out to be the following:
\begin{align}
\partial^\nu\bigg\{
\kappa p\frac{x_\mu x_\nu}{\tau^2} - \Big(p {-} \zeta\frac d\tau\Big)\cdot\Big(g_{\mu\nu} {-} \frac{x_\mu x_\nu}{\tau^2}\Big) \bigg\} = 0.
\end{align}
Performing the derivations (keeping in mind that $\zeta$ may also have a coordinate dependence) and projecting the resulting expression in the
direction of $x^\mu$ as well as pseudo-orthogonal to $x^\mu$, we arrive at the following two equations:
\begin{align}
d\Big[(\kappa{+}1)p-\zeta\frac d\tau\Big]+\kappa x^\nu\partial_\nu p &= 0,\label{e:energy:p}\\
\Big(g_{\mu\nu}-\frac{x_\mu x_\nu}{\tau^2}\Big)\Big[\frac d\tau\partial^\nu\zeta - \partial^\nu p\Big] &= 0.
\end{align}
It is easy to verify that the second one is equivalent to the condition that $p-\frac d\tau\zeta$ is a function of $\tau$ alone, which function we
now temporarily denote by $\Phi(\tau)$. In Eq.~(\ref{e:energy:p}) we can write $x^\nu\partial_\nu = \tau\partial_\tau$ to arrive at the
conditions\footnote{Here and below $\partial_\tau$ is understood in the sense that it means differentiating with respect to $\tau$ while keeping
the coordinates  pseudo-orthogonal to $\tau$ fixed. One such convenient coordinate system is e.g. the so-called spherical Rindler coordinate system:
the $\tau$ variable supplemented by $\eta$, and the two-variable $\mathbf n$ unit space-vector, so that $t = \tau\cosh\eta$, and the space-like
component of $x^\mu$ is expressed as $\mathbf r = \mathbf n\cdot \tau\sinh\eta$.}:
\begin{align}
p - \frac d\tau\zeta &= \Phi(\tau)\quad\textnormal{arbitrary}, \label{e:Phitau}\\
\kappa\partial_\tau p + \frac{d(\kappa{+}1)}{\tau}p - \frac{d^2}{\tau^2}\zeta &= 0 .\label{e:ptauzeta}
\end{align}

Turing towards Israel-Stewart hydrodynamics, the relaxation equations \eqref{e:IS1}-\eqref{e:IS2} can be written up as:
\begin{align}
\tau_{_\Pi} \partial_\tau \Pi &= -\zeta \frac{d}{\tau} - \Pi,\label{e:IS1Hubble}\\
\tau_\pi \partial_\tau \pi^{\mu\nu} &= 2\eta \sigma^{\mu\nu} - \pi^{\mu\nu},\label{e:IS2Hubble}
\end{align}
where $\partial_\tau$ denotes $u_\rho \partial^\rho$. Since $\sigma^{\mu\nu}=0$ in case of Hubble flow, the second equation is solved by $\pi^{\mu\nu}=0$ independently of $\eta$. Furthermore, the hydrodynamics equations ($\partial_\nu T^{\mu\nu}=0$) with $\Pi$ can also be written up. The equations parallel and orthogonal to $u^\mu$ yield equations similar to \eqref{e:Phitau}-\eqref{e:ptauzeta}:
\begin{align}
p + \Pi &= \Psi(\tau)\quad\textnormal{arbitrary}, \label{e:Psitau}\\
\kappa\partial_\tau p + \frac{d}{\tau}\left((\kappa+1) p + \Pi \right) &= 0 .\label{e:pPitau}
\end{align}
This is then coupled to Eq.~\eqref{e:IS1Hubble} and has to be solved based on the assumption for $\zeta$ substituted to \eqref{e:IS1Hubble}.

We restrict ourselves to the case when $p {\equiv} p(\tau)$ and $\Pi {\equiv} \Pi(\tau)$, that is, when $p$ and $\Pi$ depend only on $\tau$.
In Cases A and B, when $\zeta {=} \zeta_0 {=} $const., this is not an additional assumption but something that follows evidently
from Eq.~(\ref{e:Phitau}). Slightly less evidently, but the same is true for Case C, when $\zeta$ is not constant,
but there is no conserved $n$ particle density\footnote{In this Case C $\zeta$ depends only on $T$, which is in a one-to-one correspondence
with $p$, so $\zeta$ in turn depends only on $p$, so from Eq.~(\ref{e:Phitau}) we arrive at the necessary conclusion that $p \equiv p(\tau)$.}.
In Cases D and E, the condition that $p$ depends only on $\tau$ does not necessarily follow from the equations encountered
so far. In these cases $n$ is non-vanishing, and $\zeta$ depends on either $T$ or $n$ as specified by Eqs.~(\ref{e:cased}) or (\ref{e:casee}).
Allowing $p$ to have a more general form beyond $p(\tau)$ would make the equations so complicated that it seems hopeless
(and futile) to investigate this direction any further; in these cases we thus {\it assume} that $p \equiv p(\tau)$. But as soon as this is assumed,
in these cases now it also follows that $T$, and in turn, $n$ can be a function of $\tau$ alone.
In particular, in Cases D and E we must take $\mathcal V(S)\equiv 1$ for the up to now arbitrary
scaling function in the solution for the continuity equation, Eq.~(\ref{e:nsol}).

The problem of finding viscous solutions with the assumption of Hubble-type velocity profile has thus been reduced to the task of finding solutions for the following ordinary differential equations.
\begin{align}
&\textnormal{Navier-Stokes case:}\nonumber\\
\kappa\frac{dp}{d\tau} &= - \frac{d(\kappa{+}1)}{\tau}p + \frac{d^2}{\tau^2}\zeta.\label{e:ptaudiff}\\
&\textnormal{Israel-Stewart case:}\nonumber\\
\kappa\frac{dp}{d\tau} &= - \frac{d}{\tau}\left((\kappa+1) p + \Pi \right),\label{e:ptaudiffIS}\\
\frac{d\Pi}{d\tau} &= -\frac{\zeta}{\tau_{_\Pi}} \frac{d}{\tau} - \frac{\Pi}{\tau_{_\Pi}}.\label{e:PitaudiffIS}
\end{align}
Here $\zeta$ must be substituted as a function of thermodynamical quantities, as follows from the assumptions in the cases outlined in the previous section.
While the above equation provides a general solution (in the form of an ordinary differential equation) valid for a general $\zeta$ function, below we write up the solutions for $p$ in each of the specific cases mentioned in Eqs.~(\ref{e:casea})-(\ref{e:casee}). In all of the cases, the temperature $T$ can be written up using the expression for $p$:
\begin{align}
T &= T_0\left(\frac{p}{p_0}\right)^\frac{1}{\kappa+1}&&\textnormal{in case of no conserved $n$,}\label{e:Tsol:non}\\
T &= p/n&&\textnormal{in case of non-vanishing $n$.}\label{e:Tsol:yesn}
\end{align}

\section{Simple solutions for non-vanishing bulk viscosity}\label{s:solution}

\underline{\textit{In Cases A and B,}} when $\zeta = \zeta_0$=const., the equation is simple to solve:
\begin{align}
\kappa\frac{dp}{d\tau} +& \frac{d(\kappa{+}1)}{\tau}p - \frac{d^2}{\tau^2}\zeta_0 = 0 \qquad\Rightarrow \nonumber\\
\Rightarrow\quad p(\tau) =& \left[p_0 - \frac{d^2}{(\kappa{+}1)d - \kappa}\frac{\zeta_0}{\tau_0}\right]
\left(\frac{\tau_0}{\tau}\right)^{d\frac{\kappa{+}1}{\kappa}} + \nonumber\\ &+ \frac{d^2}{(\kappa{+}1)d - \kappa}\frac{\zeta_0}{\tau} .
\end{align}
\underline{\textit{In Case C,}} in the expression of $\zeta(T)$ we substitute $T$ as a function of $p$ to write up the equation:
\begin{align}
\kappa\frac{dp}{d\tau} + \frac{d(\kappa{+}1)}{\tau}p - \frac{d^2}{\tau^2}\zeta_0\left(\frac{p}{p_0}\right)^{\frac{\kappa}{\kappa+1}} = 0.
\end{align}
The solution of this equation when $\kappa\neq d$ is
\begin{align}
 p(\tau)
= p_0\bigg\{\left(1+\frac{d^2}{(\kappa{+}1)(\kappa{-}d)}\frac{\zeta_0}{p_0\tau_0}\right)\left(\frac{\tau_0}{\tau}\right)^{\frac d\kappa} -\nonumber\\
\quad- \frac{d^2}{(\kappa{+}1)(\kappa{-}d)}\frac{\zeta_0}{p_0}\frac 1\tau\bigg\}^{\kappa+1},
\end{align}
while in the $\kappa {=} d$ case it is
\begin{align}
 p(\tau) = p_0\bigg[1 + \frac{\kappa}{\kappa{+}1}\frac{\zeta_0}{p_0\tau_0}\ln\frac\tau{\tau_0}\bigg]\left(\frac{\tau_0}{\tau}\right)^{\kappa+1}.
\end{align}
\underline{\textit{In Case D,}} we have the equation as
\begin{align}
&\kappa\frac{dp}{d\tau} + \frac{d(\kappa{+}1)}{\tau}p - \frac{d^2}{\tau^2}\zeta_0\left(\frac{\tau_0}\tau\right)^d = 0,
\end{align}
and the solution for $\kappa{\neq}d$ as
\begin{align}
p(\tau) = \left[p_0{+}\frac{d^2}{\kappa{-}d}\frac{\zeta_0}{\tau_0}\right]\left(\frac{\tau_0}{\tau}\right)^{\frac{\kappa+1}{\kappa} d}
- \frac{d^2}{\kappa{-}d}\frac{\zeta_0}{\tau_0}\frac{\tau_0^{d+1}}{\tau^{d+1}},
\end{align}
while for $\kappa {=} d$ we have
\begin{align}
p(\tau) = p_0\left[1 + \frac{\zeta_0\kappa}{p_0\tau_0}\ln\frac{\tau}{\tau_0}\right]\left(\frac{\tau_0}{\tau}\right)^{\kappa+1} .
\end{align}
\underline{\textit{In Case E,}} in $\zeta(T)$ we substitute $T$ expressed through $p$ and $n$: $T = p/n$. To this end we invoke the solution for $n$,
Eq.~(\ref{e:nsol}) with $\mathcal V(S)=1$. We find that in this case the equation is
\begin{align}
\kappa\frac{dp}{d\tau} + \frac{d(\kappa{+}1)}{\tau}&p -
\frac{d^2}{\tau^2}\zeta_0\left(\frac{p}{p_0}\left(\frac\tau{\tau_0}\right)^d\right)^\kappa = 0,
\label{e:caseEdiff}
\end{align}
and the solution reads as
\begin{align}
p(\tau)
&= p_0\bigg\{\bigg(1-\frac{d^2(\kappa{-}1)}{\kappa{-}d}\frac{\zeta_0}{p_0\tau_0}\bigg)\left(\frac{\tau}{\tau_0}\right)^{d\frac{\kappa^2-1}{\kappa}}
+\nonumber\\
&+ \frac{d^2(\kappa{-}1)}{\kappa{-}d}\frac{\zeta_0}{p_0\tau_0}\Big(\frac{\tau}{\tau_0}\Big)^{d\kappa -1}
\bigg\}^{-\frac1{\kappa-1}}
\label{e:caseEsol}
\end{align}
for $\kappa{\neq}d$, while for $\kappa{=}d$ it can be written as
\begin{align}
p(\tau)
&= p_0\Big(\frac{\tau_0}{\tau}\Big)^{\kappa+1}\bigg\{1 - \kappa(\kappa{-}1)\frac{\zeta_0}{p_0\tau_0}\ln\frac{\tau}{\tau_0}
\bigg\}^{-\frac1{\kappa-1}}.
\label{e:caseEsolln}
\end{align}
It should be noted here that in the case of $\kappa>d$ (as it is normally assumed) as well as $\frac{d^2(\kappa{-}1)}{\kappa{-}d}\frac{\zeta_0}{p_0\tau_0}<1$ (which is fulfilled in case of a moderate $\zeta_0$ value) this solution behaves well. If these conditions are not met
(including the $\kappa{=}d$ exceptional case, written up separately above) the quantity to be raised to the $-\frac1{\kappa-1}$th power necessarily becomes zero at some $\tau$ after the start of the time evolution ($\tau_0$). Hence if the above mentioned conditions are not met, then this solution is not physical.
\underline{\textit{Finally, in Case F,}} $\zeta=\zeta_0 \Pi/\Pi_0$, hence we get from \eqref{e:PitaudiffIS}:
\begin{align}
\frac{d\Pi}{d\tau} &= -\frac{\zeta_0}{\tau_{_\Pi}}\frac{\Pi}{\Pi_0} \frac{d}{\tau} - \frac{\Pi}{\tau_{_\Pi}}
\end{align}
which yields
\begin{align}
\Pi (\tau) =\Pi_0 \left(\frac{\tau_0}{\tau}\right)^{\frac{d \zeta_0}{\Pi_0 \tau_{_\Pi} }}
 \exp\left[-\frac{\tau-\tau_0}{\tau_{_\Pi}}\right] .
\end{align}
This can be substituted back to Eq.~\eqref{e:ptaudiffIS} to obtain
\begin{align}
p(\tau) = p_A\left(\frac{\tau_0}{\tau}\right)^{d+\frac{d}{\kappa}}
 \left[1{+}\frac{p_0{-}p_A}{p_A}\frac{\Gamma\left(B,\tau/\tau_{_\Pi}\right)}{\Gamma\left(B,\tau_0/\tau_{_\Pi}\right)}\right],\label{e:caseFsol}
\end{align}
where
\begin{align}
p_A &= p_0 - \frac{\Pi_0 d}{\kappa} \Gamma \left(B,\frac{\tau }{\tau_{_\Pi} }\right)
\left(\frac{\tau_0}{\tau_{_\Pi}}\right)^{-B}e^{\tau_0/\tau_{_\Pi}}\textnormal{ and}\\
B &= d+\frac{d}{\kappa }-\frac{\zeta_0 d}{\Pi_0 \tau_{_\Pi} }
\end{align}
It should be noted here that two new parameters are introduced in this case: $\Pi_0$ (independently setting the bulk pressure scale) and $\tau_{_\Pi}$ (setting the relaxation scale). Furthermore $\Pi_0<0$ is required~\cite{Csorgo:2020iug} in order to fulfill the second law of thermodynamics (i.e. that there is entropy production). This exact solution belongs to the class of asymptotically perfect fluid solutions of dissipative relativistic hydrodynamics detailed in Ref.~\cite{Csorgo:2020iug}

Table~\ref{t:cases} summarizes the general characteristics of the solutions found so far.

\begin{table}%[H]
\begin{center}
\begin{tabular}{|c|c|c|c|c|}
\hline\hline
Case & $\zeta$ & $T$ definition & asymptotics \\
\hline\hline
 (A)  & $\zeta_0\textnormal{ (const)}$ & $p = p_0(T/T_0)^{\kappa+1}$ & physical           \\ \hline
 (B)  & $\zeta_0\textnormal{ (const)}$ & $p = nT$                    & $T(\tau)\to\infty$ \\ \hline
 (C)  & $\zeta_0(T/T_0)^\kappa$        & $p = p_0(T/T_0)^{\kappa+1}$ & physical           \\ \hline
 (D)  & $\zeta_0(n/n_0)$               & $p = nT$                    & physical           \\ \hline
 (E)  & $\zeta_0(T/T_0)^\kappa$        & $p = nT$                    & \makecell[cc]{conditionally\\physical}\\
 (F)  & $\zeta_0(\Pi/\Pi_0)$           & $p = p_0(T/T_0)^{\kappa+1}$ & physical           \\ \hline
  \hline
\end{tabular}
\caption{New exact viscous hydrodynamical solutions are summarized in this table. Here ``physical'' refers to all thermodynamical quantities decreasing to zero for $\tau\to\infty$. Case E is physical only if $\kappa>d$ and $\frac{d^2(\kappa{-}1)}{\kappa{-}d}\frac{\zeta_0}{p_0\tau_0}<1$, see more details in the main text after Eq.~\eqref{e:caseEsolln}.}\label{t:cases}
\end{center}
\end{table}

Let us close this section with some general remarks.
When solving the first order differential equation (\ref{e:ptaudiff}) for the $p(\tau)$ dependence in the different cases, a constant of integration appears.
In our previous treatment, this constant was always chosen in a way that the $p_0$ value (appearing in all expressions of the $p(\tau)$ dependence) has
the simple meaning that at $\tau = \tau_0$, the pressure takes the value $p(\tau_0) = p_0$. Also, the notation of $\zeta_0$ was chosen in a way so that
at the beginning of the time evolution, $\tau {=} \tau_0$, the value of the bulk viscosity is $\zeta_0$ in all cases, irrespective of whether this is an
assumed constant value throughout the evolution (as in Cases A and B) or if it changes over time (as in Cases C, D, E).

As hinted at before, the role of the thermal conductivity $\lambda$ can be investigated straightforwardly in all Navier-Stokes cases. If (and in our framework, only
if) the temperature $T$ depends only on $\tau$, we see from the expression of the thermal conductive part of the $T_{\mu\nu}$ energy-momentum tensor,
Eq.~(\ref{e:Tmunu:expr}), that all the terms containing $\lambda$ cancel. (This is a special feature of the simple Hubble-like velocity field.) So the
conclusion is that if (and only if) $T\equiv T(\tau)$, all of our previous solutions remain valid even with arbitrary thermal conductivity
terms, ie.\@ for any arbitrary $\lambda\neq0$.

The fulfillment of the condition $T\equiv T(\tau)$ is not some far-reaching further specification, but rather fits very naturally to the solutions
presented above. We have seen in the paragraphs after Eq.~(\ref{e:ptauzeta}) towards the end of Sec.~\ref{s:searching} that $p\equiv p(\tau)$,
ie.\@ $p$ depends only on $\tau$ (and actually found its expression in all cases). In Cases A and C, there is no conserved density $n$, and there
is a one-to-one correspondence between $p$ and $T$, so $p\equiv p(\tau)$ automatically leads to $T\equiv T(\tau)$. In the other cases we have
$p = nT$, and the solution for $n$, Eq.~(\ref{e:nsol}), contains an arbitrary $\mathcal V(S)$ function of the scaling variable $S$.
From this and Eq.~(\ref{e:Tsol:yesn}) we thus see that in these cases (B, D, E) the expression of $T$ will contain $1/\mathcal V(S)$.
So in Cases B, D, and E, the fulfillment of the $T\equiv T(\tau)$ condition requires that we set $\mathcal V(S)\equiv 1$: with this, the
solutions will be valid for any type of thermal conductivity.

Given that the Hubble flow profile has no shear, the solutions given above are valid for arbitrary shear viscosity coefficients. Even if the shear viscosity coefficient (or the kinematic viscosity, $\eta/s$) has any dependence on the temperature $T$ or on the number density $n$, the shear viscosity effects totally cancel from these solutions. So the solutions presented above are valid not only for any type of heat conductivity, but also for any type of shear viscosity as well.

Finally let us note, that the above solutions immediately show, that two of the above cases lead to unphysical results. In Case B, while the pressure asymptotically decreases to zero as $p\propto\tau^{-1}$, the temperature diverges if $d>1$, as $p=nT$ and $n\propto\tau^{-d}$, hence asymptotically $T\propto\tau^{d-1}$. An ever increasing temperature is however unphysical generally, in particular it is not a realistic feature for the QGP observed in relativistic heavy ion collisions. Furthermore, in Case E, expression (\ref{e:caseEsol}) for any conceivable $\zeta_0$, $p_0$ and $\tau_0$ values, for $d=3$ and $\kappa\ge1{/}2$ (which is a
quite physical condition) already the pressure asymptotically increases with $\tau$. This too is quite unphysical. The reason for this might be explained by noting that the (\ref{e:caseEdiff}) differential equation contains an explicit, $\tau$-dependent,
strongly increasing ,,source term''. The increase of this last term in turn follows from the assumption made here, namely that the bulk viscosity is
proportional to $T^\kappa$, but $T$ must be calculated by dividing $p$ by $n$, and $n$ decreases as $\propto\tau^{-d}$. Taking these together, it is
indeed plausible that in this case $p$ as well as $T$ diverges for $\tau\to\infty$, which reflects the fact that the assumption of Case E itself is
not entirely physical.

\section{Illustration and discussion}\label{s:illustration}

\begin{figure}
\includegraphics[width=0.9\linewidth]{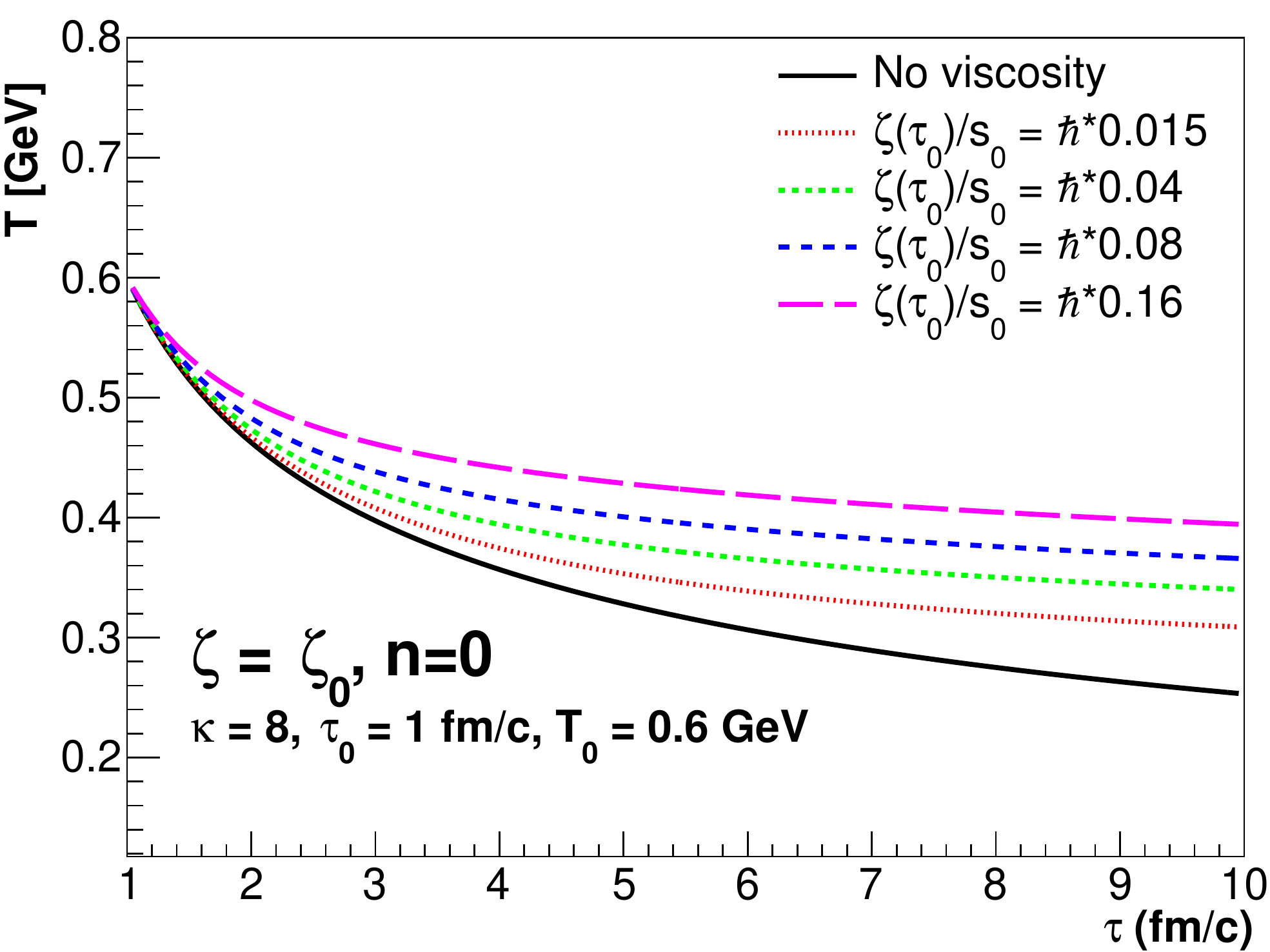}
\caption{(Color online) Temperature evolution in Case A: no conserved $n$ and $\zeta=\zeta_0=$const.}
\label{f:caseA:energy}
\end{figure}

\begin{figure}
\includegraphics[width=0.9\linewidth]{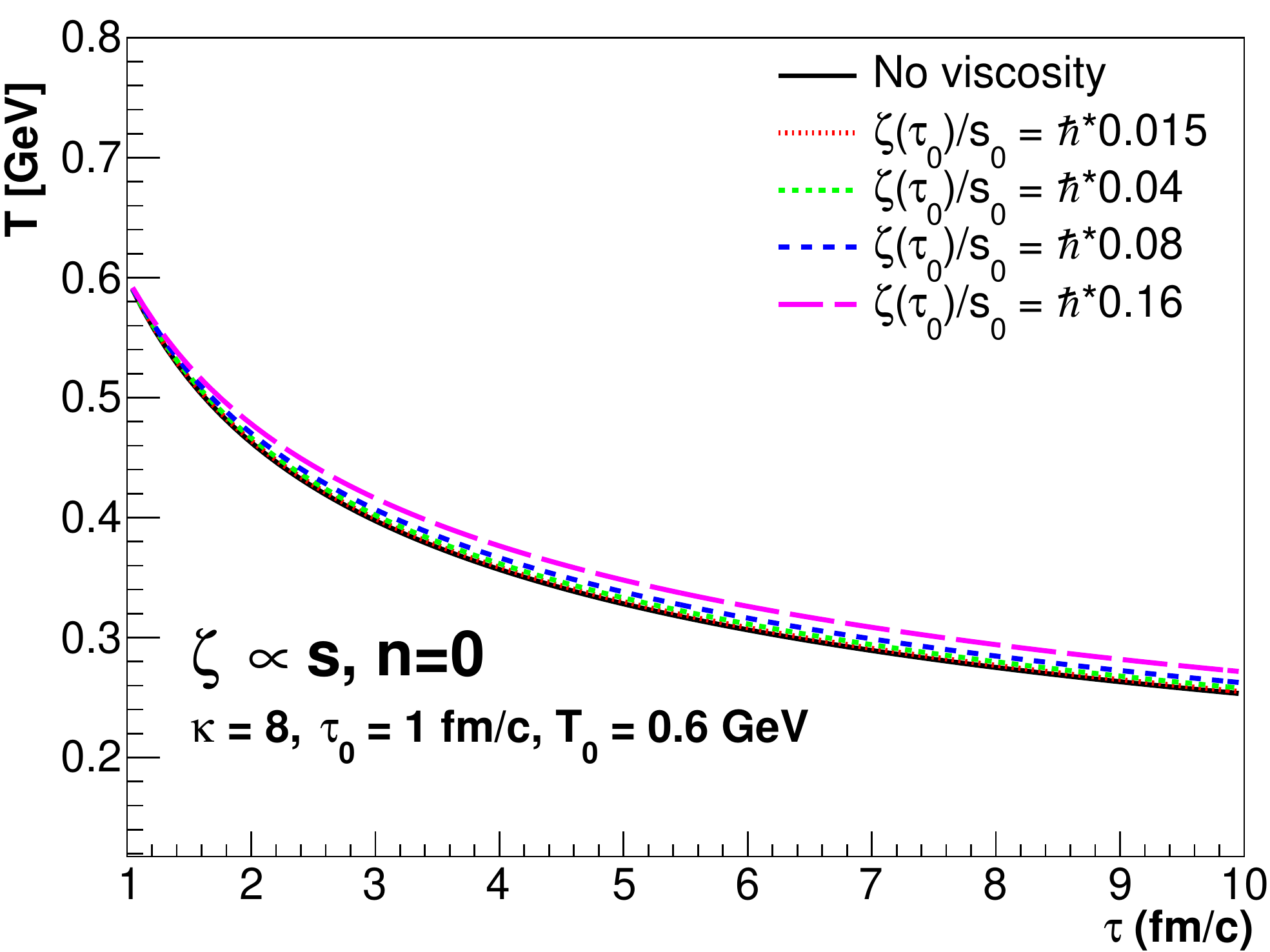}
\caption{(Color online) Temperature evolution in Case C: no conserved $n$, and $\zeta=\zeta_0(s/s_0)$, entropy dependent bulk viscosity.}
\label{f:caseC:energy}
\end{figure}

\begin{figure}
\includegraphics[width=0.9\linewidth]{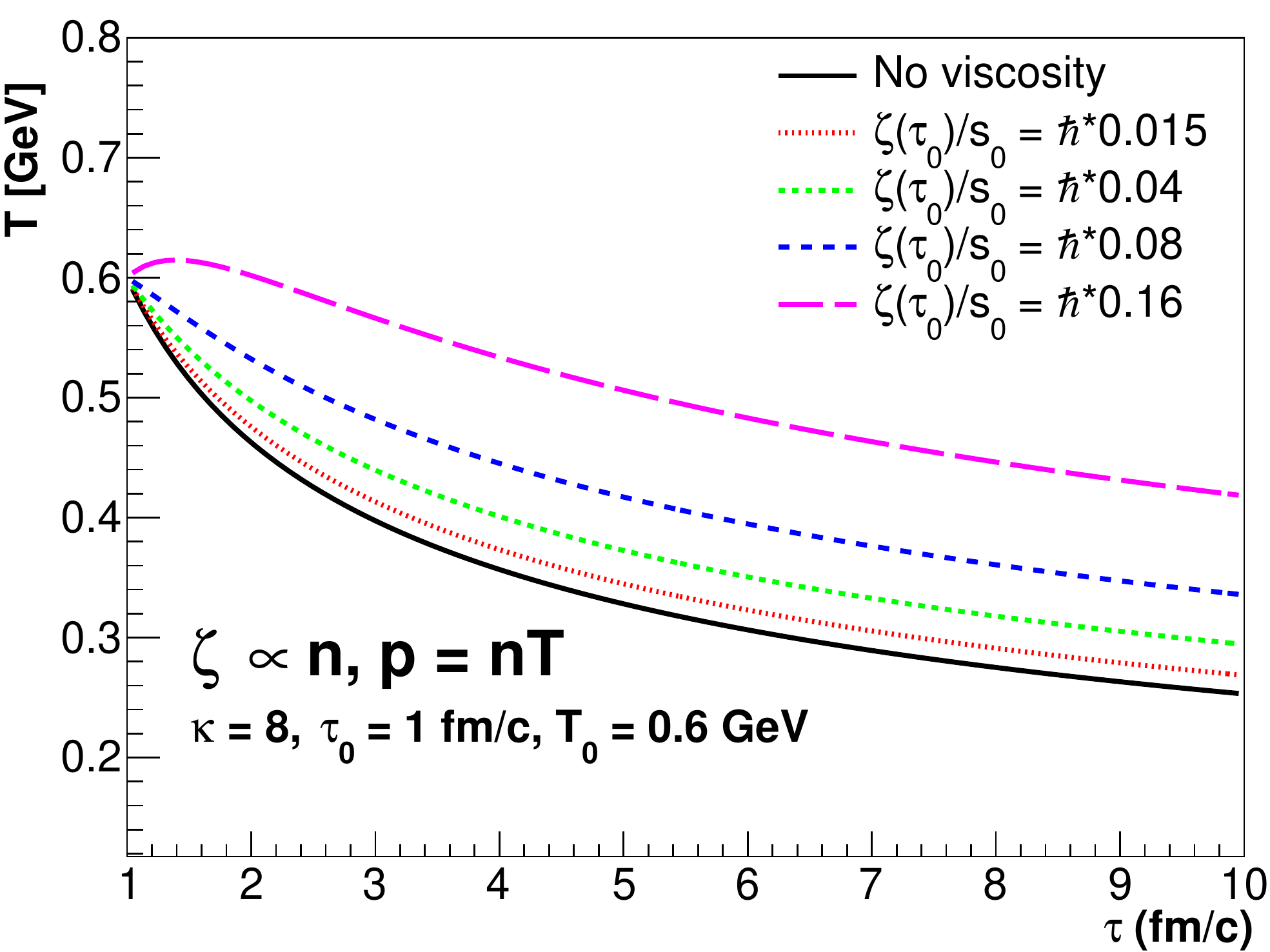}
\caption{(Color online) Temperature evolution in Case D: non-vanishing conserved $n$, and $\zeta=\zeta_0(n/n_0)$, proportional to $n$.}
\label{f:caseD:energy}
\end{figure}

It is interesting to plot the time evolution of the temperature as a function of proper time. In Figs.~\ref{f:caseA:energy}--\ref{f:caseD:energy} we
see this for the physically relevant cases (Cases A, C and D). In these plots, we vary the value of the bulk viscosity $\zeta_0$ (the constant value or the
initial value, depending on the assumed behavior of $\zeta$). Furthermore, for comparison we also show the time evolution corresponding to the case
of no viscosity (which was an earlier known result for the Hubble flow and serves as a benchmark to investigate the effects of bulk viscosity).

Of course the five scenarios laid out so far, Cases A through E, as well as the solutions themselves, are fundamentally different from each other.
Indeed it turns out that in some of the cases, the temperature evolution behaves rather unphysically, while in other cases the effects of bulk viscosity
are quite moderate. One reason for this difference is that the same choice for $\zeta_0$ has completely different meanings in the different cases.
The reason for the divergence of $T$ in Case E was explained after Eq.~(\ref{e:caseEsol}). The reason for the increase of $T$ in Case B is a subtler
one. In this case for $\tau{\to}\infty$ the pressure $p$ tends to zero, however, the conserved density $n$ decreases faster.
This leads to $T = p/n$ diverging. Due to this, we refrain from plotting the temperature evolution in Cases B and E.

\begin{figure}
\includegraphics[width=0.9\linewidth]{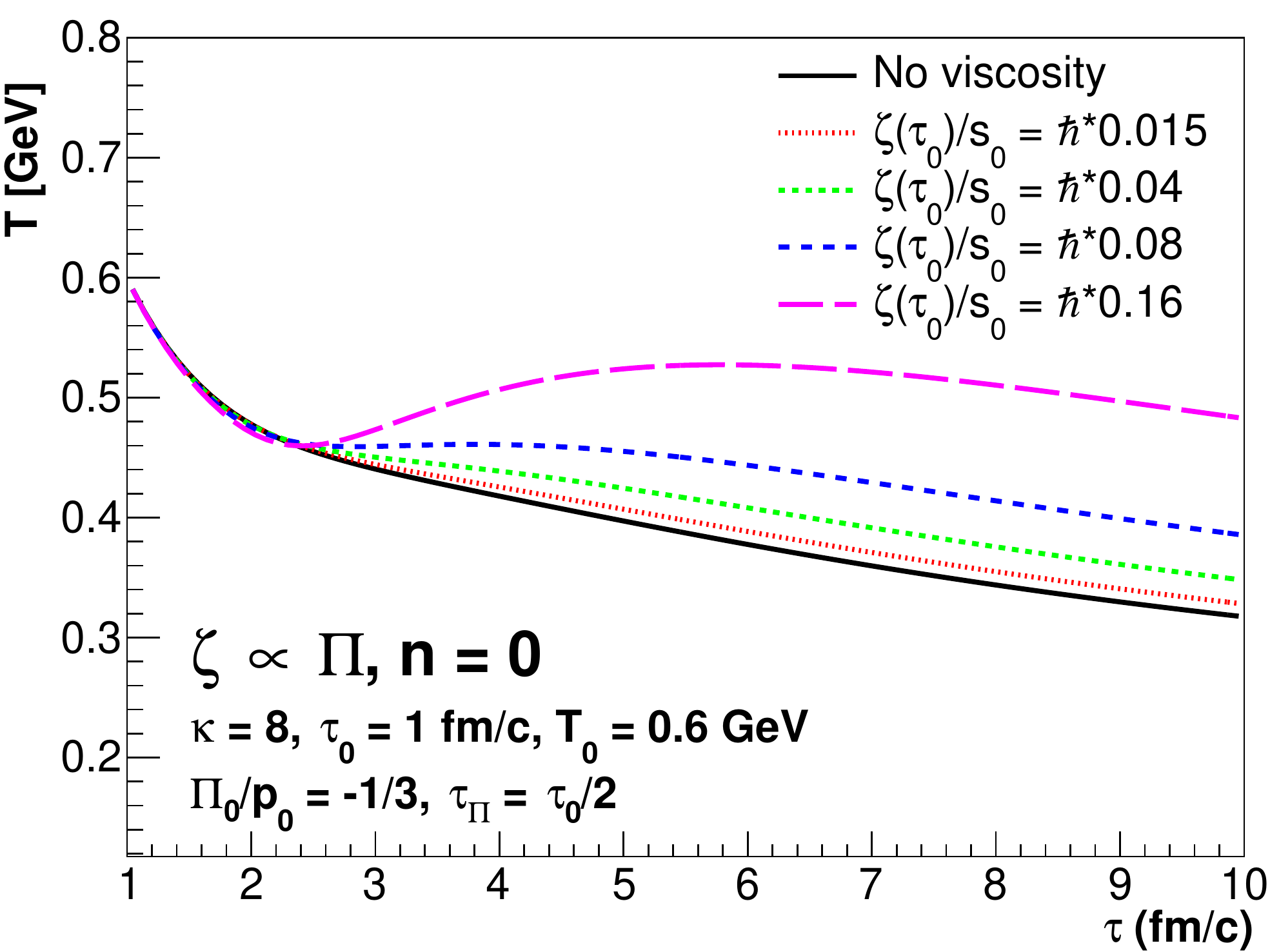}
\caption{(Color online) Temperature evolution in Case F: no conserved $n$, and $\zeta=\zeta_0(\Pi/\Pi_0)$, proportional to $\Pi$.}
\label{f:caseF:energy}
\end{figure}

\section{Summary}

We have presented a new family of exact analytic solutions of first-order viscous relativistic hydrodynamics. Utilizing a simple Hubble-like velocity
profile and ellipsoidal temperature and density profile, we have obtained a variety of solutions under different assumptions on the overall behavior
of the bulk viscosity of the fluid. A very interesting feature of our solutions is that the effect
of the shear viscosity completely cancels; put in another way, our solutions remain valid and do not change in the slightest for any conceivable
assumption for the shear viscosity. In this way, our solutions based on the Hubble-like velocity profile provide excellent opportunity to study
the effects of bulk viscosity (irrespective of shear viscosity). These solutions also provide an excellent testing opportunity for the numerical solutions of relativistic viscous hydrodynamics,
where the effects of numerical viscosity can be cross-checked against the exact solution presented in this paper, valid for arbitrary kinematic shear viscosity coefficients or functions. This is important because the theoretical understanding and modeling of bulk
viscosity of the strongly interacting matter produced in heavy ion collisions is far less developed than that of shear viscosity. Concerning
thermal conductivity (the remaining dissipative coefficient in a first-order theory), in the case when spherical symmetry is retained,
the presented solutions also work with any assumed value of it (i.e. its effect also cancels).

The theoretical uncertainty of bulk viscosity modeling manifests itself not only in the predicted concrete numerical values but also in the general trend
of the time evolution of the bulk viscosity (or rather, the dependence of it on other thermodynamic parameters such as phase transition type, pressure,
temperature, etc). The different ,,scenarios'' for the evolution of bulk viscosity lead to qualitatively different time evolution of the energy density
and temperature of the expanding system. In our work we have treated five different plausible scenarios for the bulk viscosity and other thermodynamic
properties (i.e. equation of state) of the system, and examined the effect that bulk viscosity has on time evolution. It indeed turned out that
they are very much different; in some cases the bulk viscosity has minimal effect, in other cases even the smallest bulk viscosity leads to (unphysical)
re-heating of the system. The considered special cases A, B, C, D and E represent specific choices for the dependence of the bulk viscosity coefficient on the local properties of the matter. They correspond to the first known examples of exact solutions of relativistic fireball hydrodynamics with bulk and shear viscosity. One may expect that they all correspond to special cases of a more general solution. Finding such a generalized solution, for example for a temperature dependent speed of sound and/or kinematic viscosity coefficients is one of our current reseach directions.

The work written up in this paper, among the first ones of such type according to our knowledge, represents a first step towards unveiling analytic
solutions of viscous hydrodynamics. The Hubble-type velocity profile utilized here greatly simplifies the problem at hand. Although it is conceivable that
a Hubble-like flow profile fits naturally to the description of heavy-ion collisions, it is obvious that a next research direction can be the generalization
to other, more complicated velocity profiles as well as more complicated equations of state. Also it would be worthwhile to consider second order theories
of viscosity (such as Israel-Stewart theory). These are only a handful of directions in which our presented work can be developed in the future.

\section*{Acknowledgements}
The authors would like to thank G\'abor Kasza for the careful reading of our manuscript and insightful discussions. Our research has been partially supported by the COST Action CA15213 of the European Union, the Hungarian NKIFH grants No. FK-123842 and FK-123959, the Hungarian EFOP 3.6.1-16-2016-00001 project. M. Csan\'ad and M. Nagy were supported by the J\'anos Bolyai Research Scholarship and the \'UNKP-19-4 New National Excellence Program of the Ministry for Innovation and Technology

\end{document}